\documentclass[12pt]{iopart}
\usepackage{graphicx}

\begin{document}

\title{Langevin equation with super-heavy-tailed noise}

\author{S I Denisov$^{1,2}$, H Kantz$^{1}$ and P H\"{a}nggi$^3$}

\address{$^1$Max-Planck-Institut f\"{u}r Physik komplexer Systeme,
N\"{o}thnitzer Stra{\ss}e 38, D-01187 Dresden, Germany\\
$^2$Sumy State University, 2 Rimsky-Korsakov Street, UA-40007 Sumy, Ukraine\\
$^3$Institut f\"{u}r Physik, Universit\"{a}t Augsburg,
Universit\"{a}tsstra{\ss}e 1, D-86135 Augsburg, Germany}
\ead{stdenis@pks.mpg.de}

\begin{abstract}
We extend the Langevin approach to a class of driving noises whose generating
processes have independent increments with super-heavy-tailed distributions.
The time-dependent generalized Fokker-Planck equation that corres\-ponds to the
first-order Langevin equation driven by such a noise is derived and solved
exactly. This noise generates two probabilistic states of the system, survived
and absorbed, that are equivalent to those for a classical particle in an
absorbing medium. The connection between the rate of absorption and the
super-heavy-tailed distribution of the increments is established analytically.
A numerical scheme for the simulation of the Langevin equation with
super-heavy-tailed noise is developed and used to verify our theoretical
results.
\end{abstract}



\maketitle

\section{Introduction}
\label{Intr}

At the beginning of the last century, Paul Langevin, a prominent French
physicist, proposed an alternative description of Brownian motion \cite{Lang}.
In contrast to earlier Einstein's approach \cite{Ein}, which deals with the
Fokker-Planck equation for the proba\-bility density of an ensemble of
independent Brownian particles, Langevin used the equation of motion for a
single Brownian particle, the so-called Langevin equation. In fact, applying
Newton's second law with a random force arising from the surrounding medium, he
introduced a first example of stochastic differential equations. Now the
Langevin approach is one of the most effective and widely used tools for
studying the effects of a fluctuating environment \cite{CKW}.

In physics, many other stochastic equations that account for the influence of
the environment are also often termed ``Langevin equations". Among them, the
dimensionless first-order (overdamped) Langevin equation
\begin{equation}
    \dot{x}(t) = f(x(t),t) + \xi(t)
\label{Langevin}
\end{equation}
is the simplest and at the same time most important. It describes  noise
effects in very different systems, but to be specific we will call $x(t)$ a
particle position [$x(0)=0$], $f(x,t)$ a deterministic force field, and
$\xi(t)$ a random force (noise) resulting from a fluctuating environment. The
Langevin approach is particularly useful when the noise $\xi(t)$ can be
approximated by the time derivative of the noise generating process $\eta(t)$,
i.e., random process with stationary and independent increments $\Delta \eta(t)
= \eta(t+\tau) - \eta(t)$. The reason is that in this case the solution of
equation (\ref{Langevin}) is a Markov process whose future behavior is
determined only by the present state, see e.g. \cite{HT,HL}.

The main statistical characteristic of the particle position $x(t)$, its
probability density $P(x,t)$, depends fundamentally on the probability density
$p(\Delta \eta, \tau)$ of the increments $\Delta \eta(t)$ for a sufficiently
small time increment $\tau$. On a phenomenological level, $p(\Delta \eta,
\tau)$ must be chosen to obtain the best representation of the environmental
fluctuations. If these fluctuations are the result of many independent random
variables with zero mean and finite variance, then, according to the central
limit theorem \cite{Fel}, $p(\Delta \eta, \tau)$ can be approximated by a
Gaussian probability density with zero mean and a variance proportional to
$\tau$. Accordingly, $\eta(t)$ becomes a Wiener process and $\xi(t)$ reduces to
Gaussian white noise. As is well known, in this case the process $x(t)$ is
continuous and $P(x,t)$ satisfies the ordinary Fokker-Planck equation
\cite{CKW}--\cite{R} which describes a huge variety of noise phenomena. If the
above mentioned random variables have no finite variance then, as the
generalized central limit theorem suggests \cite{Fel}, $p(\Delta \eta, \tau)$
can be chosen in the form of a L\'{e}vy stable probability density. This choice
of $p(\Delta \eta, \tau)$ corresponds to L\'{e}vy stable noise $\xi(t)$ and
leads to the fractional Fokker-Planck equation for $P(x,t)$
\cite{JMF}--\cite{DS}. The Langevin equation (\ref{Langevin}) driven by
L\'{e}vy stable noise and the corresponding fractional Fokker-Planck equation
describe the so-called L\'{e}vy flights, i.e., random processes exhibiting rare
but large jumps \cite{SZF}--\cite{DSU}.

Recently, it has been shown \cite{DHK} that the ordinary and fractional
Fokker-Planck equations hold also for the probability densities represented in
the form $p(\Delta \eta, \tau) = p(\Delta \eta/a(\tau)) /a(\tau)$, where
$a(\tau)$ is a scale function that tends to zero as $\tau \to 0$, and $p(y)$ is
a probability density satisfying the condition $\lim_{\epsilon \to 0}
p(y/\epsilon)/\epsilon = \delta(y)$ with $\delta(y)$ being the Dirac delta
function. Specifically, the ordinary equation holds for a class of probability
densities $p(y)$ with finite variance, and the fractional one for a class of
heavy-tailed $p(y)$, i.e., probability densities with power-law tails and
infinite second moment. Although these classes cover the most important cases,
they do not exhaust all possible probability densities: There also exists a
class of super-heavy-tailed probability densities $p(y)$ which have never been
used in the Langevin approach. According to the definition, these densities
decay so slowly that all fractional moments $M_{\gamma}= \int_{- \infty}
^{\infty}dy\, |y|^{\gamma} p(y)$ with $\gamma>0$ are infinite (at $\gamma=0$
the normalization of $p(y)$ implies $M_{0}=1$). Because of this ``unphysical"
property, the super-heavy-tailed distributions, as well as the heavy-tailed
ones, are not directly related to ``physical" random variables having finite
variance. Nevertheless, they are often used to model some specific properties
of physical systems. Moreover, the heavy-tailed distributions have become a
common tool for studying anomalous physical phenomena, see e.g.
\cite{SZF}--\cite{CGKM} and \cite{BG}--\cite{Z}. Although the
super-heavy-tailed distributions have not yet received much attention, the
examples of extremely slow diffusion \cite{HW,DK} demonstrate the usefulness of
these distributions for describing systems with highly anomalous behavior.

The purpose of this paper is twofold. First, we wish to extend the Langevin
equation method to a class of super-heavy-tailed noises, i.e. noises arising
from the super-heavy-tailed distributions. The importance of this problem is
that these noises, together with the previously known, exhaust all possible
noises associated with the probability density $p(y)$ and, as a consequence,
enhance the capability of this method. Second, we intend to show that
super-heavy-tailed noises generate two probabilistic states of a particle,
survived and absorbed. The existence of these states makes it possible to model
the particle dynamics, which is randomly interrupted by the transition of a
particle to a qualitatively new state, within the Langevin approach. This
transition can occur in a number of physical processes, including the processes
of annihilation, evaporation and absorption. In this paper we demonstrate the
applicability of the Langevin equation (\ref{Langevin}) driven by
super-heavy-tailed noise for describing the particle dynamics, both
deterministic and random, in an absorbing medium.

\section{General results}
\label{sec:Gen}

We start our analysis with the generalized Fokker-Planck equation \cite{DHH1}
\begin{equation}
    \frac{\partial}{\partial t}P(x,t) = -\frac{\partial}{\partial x}
    f(x,t)P(x,t) + \mathcal{F}^{-1}\{P_{k}(t) \ln S_{k}\}
    \label{FP1}
\end{equation}
which corresponds to the Langevin equation (\ref{Langevin}). Here $\mathcal{F}
\{u(x)\} = u_{k} = \int_{-\infty}^{\infty} dx e^{-ikx} u(x)$ and $\mathcal
{F}^{-1}\{u_{k}\} = u(x) = (1/2\pi) \int_{-\infty}^{\infty} dk\, e^{ikx} u_{k}$
denote the direct and inverse Fourier transforms, respectively, $P_{k}(t) =
\int_{-\infty}^{\infty} dx e^{-ikx} P(x,t)$ is the characteristic function of
the particle position $x(t)$, and $S_{k}$ is the characteristic function of the
noise generating process $\eta(t)$ at $t=1$. It is assumed that the solution of
equation (\ref{FP1}) satisfies the initial, $P(x,0) = \delta(x)$, and
normalization, $\int_{-\infty}^{\infty} dx P(x,t) =1$, conditions. The
characteristic function $S_{k}$ depends on the Fourier transform of the
transition probability density $p(\Delta \eta, \tau)$ over the variable $\Delta
\eta$. In particular, if $p(\Delta \eta, \tau) = p(\Delta \eta/a(\tau))
/a(\tau)$ then
\begin{equation}
    \ln S_{k} = \lim_{\tau \to 0} \frac{1}{\tau}[p_{ka(\tau)} - 1]
    \label{S}
\end{equation}
with $p_{ka(\tau)} = \int_{-\infty}^{\infty} dy\, e^{-ika(\tau) y} p(y)$.
Equation (\ref{FP1}) reproduces all known forms of the Fokker-Planck equation
associated with the Langevin equation (\ref{Langevin}) and is valid for all
probability densities $p(y)$ \cite{DHH2}. It should be noted that because the
scale function $a(\tau)$ tends to zero as $\tau \to 0$, the dependence of $\ln
S_{k}$ on $k$ is determined by the asymptotic behavior of $p(y)$ at $y \to \pm
\infty$.

Next we consider a class of symmetric super-heavy-tailed probability densities
$p(y)$ whose asymptotic behavior at $y \to \infty$ is given by
\begin{equation}
    p(y) \sim \frac{1}{y}h(y),
    \label{as p}
\end{equation}
where $h(y)$ is a slowly varying positive function. The term ``slowly varying"
means that the asymptotic relation $h(\mu y) \sim h(y)$, i.e. $\lim_{y\to
\infty} h(\mu y)/h(y) =1$, holds for all $\mu>0$. Since the probability density
$p(y)$ is normalized, the function $h(y)$ must tend to zero more rapidly than
$1/\ln y$. On the other hand, since $y^{\nu} h(y) \to \infty$ as $y\to\infty$
for all $\nu>0$ \cite{BGT}, $h(y)$ must decay more slowly than any positive
power of $1/y$, yielding $M_{\gamma}= \infty$ ($\gamma>0$). An illustrative
example of the probability density $p(y)$ satisfying these conditions is given
in equation (\ref{p2}) below.

In order to find $\ln S_{k}$ for the examined class of probability densities,
we first rewrite (\ref{S}) in the form $\ln S_{k} =-2\lim_{\tau \to 0} Y(1/|k|
a(\tau))/\tau$ with
\begin{eqnarray}
    Y(\lambda) &=& \int_{0}^{\infty} dy [1 - \cos(y/\lambda)]p(y)
    \nonumber\\
    &=&\int_{0}^{\infty}dx\sin x \int_{\lambda x}^{\infty}dy p(y)
    \label{Y1}
\end{eqnarray}
Using either of these representations and the asymptotic formula (\ref{as p}),
one can easily show that the function $Y(\lambda)$, which according to
(\ref{Y1}) tends to zero as $\lambda \to \infty$, is slowly varying, i.e.
$Y(1/|k|a(\tau)) \sim Y(1/a (\tau))$ if $k\neq 0$ and $\tau \to 0$. Finally,
taking into account that $\ln S_{0} =0$, we obtain the general form of $\ln
S_{k}$ in the case of super-heavy-tailed $p(y)$:
\begin{equation}
    \ln S_{k} = -q(1-\delta_{k0}),
    \label{S2}
\end{equation}
where $\delta_{k0}$ is the Kronecker delta which equals $1$ if $k = 0$ and $0$
otherwise, and
\begin{equation}
    q = \lim_{\tau \to 0}\frac{2}{\tau}\,Y\bigg( \frac{1}{a(\tau)}\bigg).
    \label{q1}
\end{equation}
This limit depends on both the asymptotic behavior of $Y(\lambda)$ at $\lambda
\to \infty$ and the behavior of the scale function $a(\tau)$ at $\tau \to 0$.
Since the former is controlled by a given function $h(y)$, the nontrivial
action of super-heavy-tailed noise, which is characterized by the condition
$0<q<\infty$, occurs only at an appropriate choice of $a(\tau)$. We note that
while $a(\tau) \propto \tau^{1/2}$ for all $p(y)$ with finite variance and
$a(\tau) \propto \tau^{1/\alpha}$ almost for all heavy-tailed $p(y)$, where
$\alpha$ is an index of stability associated with the corresponding $p(y)$
\cite{DHK}, there is no universal dependence of $a(\tau)$ on $\tau$ for
different super-heavy-tailed $p(y)$. Moreover, because of the function
$Y(\lambda)$ is slowly varying, the scale function $a(\tau)$ is in general not
unique even for given $p(y)$ and $q$. We stress, however, that according to
(\ref{S2}) the influence of super-heavy-tailed noises is fully accounted by the
parameter $q$. Therefore, this lack of uniqueness is of no importance and both
the Langevin and Fokker-Planck equations remain well defined (see also the next
section).

To avoid possible problems with the interpretation of the Fokker-Planck
equation (\ref{FP1}) at the condition (\ref{S2}), let us temporarily replace
the Kronecker delta by a less singular function $\Delta_{k}$ ($\Delta_{k} = 1$
if $|k|\leq \kappa$ and $\Delta_{k} = 0$ if $|k|>\kappa$) which reduces to
$\delta_{k0}$ in the limit $\kappa\to 0$. In this case, taking $\ln S_{k} =
-q(1-\Delta_{k})$, we can write the solution of equation (\ref{FP1}) in the
form $P(x,t) = \mathcal{P} (x,t) + \mathcal{A}(x,t)$, where the terms
$\mathcal{P}(x,t)$ and $\mathcal{A} (x,t)$ are governed by the equations
\begin{eqnarray}
    &\displaystyle \frac{\partial}{\partial t}\mathcal{P}(x,t) =
    -\frac{\partial}{\partial x}f(x,t)\mathcal{P}(x,t) -
    q\mathcal{P}(x,t),&
    \label{FP2}
    \\
    &\displaystyle \frac{\partial}{\partial t}\mathcal{A}(x,t) =
    -\frac{\partial}{\partial x}f(x,t)\mathcal{A}(x,t) -
    q\mathcal{A}(x,t) + q\mathcal{F}^{-1} \{P_{k}(t)\Delta_{k}\} &
    \label{A}
\end{eqnarray}
with the initial conditions $\mathcal{P}(x,0) = \delta(x)$ and $\mathcal{A}
(x,0)=0$ and the normalization condition $P_{0}(t) = \mathcal{P}_{0}(t) +
\mathcal{A}_{0}(t) =1$. Assuming that $z(t)$ is the solution of the
deterministic equation $\dot{z}(t) = f(z(t),t)$ satisfying the initial
condition $z(0)=0$, the solution of equation (\ref{FP2}) can be written as
\begin{equation}
    \mathcal{P}(x,t) = e^{-qt}\delta[x-z(t)].
    \label{sol1}
\end{equation}
By integrating $\mathcal{P}(x,t)$ over all $x$ one finds that this solution is
not normalized to unity: $\mathcal{P}_{0}(t) = e^{-qt}$. Using this result and
the relation $\mathcal{P}_{0}(t) + \mathcal{A}_{0}(t) =1$ we obtain $\mathcal
{A}_{0}(t) = 1- e^{-qt}$. It should be noted that the function $\mathcal
{P}(x,t)$ and the normalization condition for $\mathcal{A}(x,t)$, $\mathcal
{A}_{0}(t) = 1- e^{-qt}$, do not depend on $\kappa$, i.e. they hold also at
$\kappa \to 0$ when $\Delta_{k} = \delta _{k0}$.

In contrast, it is clear from the equation
\begin{eqnarray}
    \frac{\partial}{\partial t}\mathcal{A}(x,t) &=& \!
    -\frac{\partial}{\partial x}f(x,t)\mathcal{A}(x,t) -
    q\mathcal{A}(x,t) + \frac{q\sin \kappa[x-z(t)]}
    {\pi[x-z(t)]}e^{-qt}
    \nonumber\\
    &&\! + \frac{q}{\pi} \int_{-\infty}^{\infty} dy
    \frac{\sin \kappa(x-y)} {x-y}\mathcal{A}(y,t)
    \label{A1}
\end{eqnarray}
($\mathcal{A} (x,0)=0$), which follows from (\ref{A}), (\ref{sol1}) and the
relation
\begin{equation}
    \mathcal{F}^{-1} \{P_{k}(t)\Delta_{k}\} =
    \frac{1}{2\pi}\int_{-\kappa}^{\kappa} dk\, e^{ikx}
    [\mathcal{P}_{k}(t) + \mathcal{A}_{k}(t)],
    \label{rel1}
\end{equation}
that the function $\mathcal{A}(x,t)$ depends on $\kappa$. It is not difficult
to verify that equation (\ref{A1}) provides a correct normalization for
$\mathcal{A}(x,t)$. Indeed, the integration of both sides of this equation over
all $x$ and the use of the formula $\int_{-\infty} ^{\infty} dx \sin (\kappa
x)/x = \pi$ lead to the equation $d \mathcal{A}_{0}(t)/dt = q e^{-qt}$ whose
solution satisfying the initial condition $\mathcal{A}_{0}(0) =0$ reproduces
the above result: $\mathcal {A}_{0}(t) = 1- e^{-qt}$. We note, however, that
because of replacing $\delta_{k0}$ by $\Delta_{k}$, equation (\ref{A1}) with a
fixed $\kappa$ describes the influence of super-heavy-tailed noises in an
approximate way. Moreover, this approximation is valid if $|x- z(t)|$ is much
less than the characteristic length scale $1/\kappa$ (in the opposite case the
behavior of $\mathcal{A} (x,t)$ may be unphysical with $\mathcal {A}(x,t)<0$
for some $x$ and $t$).

Assuming that $|x -z(t)| \ll 1/\kappa$ and $\kappa$ is so small that
$\mathcal{P}_{k}(t) + \mathcal{A}_{k}(t)\approx 1$, we can reduce equation
(\ref{A1}) to the form
\begin{equation}
    \frac{\partial}{\partial t}\mathcal{A}(x,t) =
    -\frac{\partial}{\partial x}f(x,t)\mathcal{A}(x,t) -
    q\mathcal{A}(x,t) + \frac{q\kappa}{\pi}.
    \label{A2}
\end{equation}
As is well known (see e.g. \cite{PZM}), its general solution is given by the
sum of a particular solution of this equation, which is proportional to
$\kappa$, and the general solution of the related homogeneous equation (when
$\kappa=0$). Since $\mathcal{A} (x,0)=0$, the latter is also proved to be
proportional to $\kappa$. Thus, the function $\mathcal{A} (x,t)$ at $|x -z(t)|
\ll 1/\kappa$ tends to zero linearly with $\kappa$. In particular, if the
external force depends only on time, i.e. $f(x,t) = f(t)$, then $z(t) =
\int_{0}^{t} dt'f(t')$ and the general solution of equation (\ref{A2}) is given
by $\mathcal{A}(x,t) = e^{-qt}\, \Phi(x - z(t)) + \kappa/\pi$, where $\Phi(x)$
is an arbitrary function. This function can be determined from the initial
condition $\mathcal{A} (x,0)=0$, $\Phi(x) = -\kappa/\pi$, yielding $\mathcal{A}
(x,t) = \kappa (1- e^{-qt})/\pi$ as $\kappa \to 0$.

The above results show that under the action of super-heavy-tailed noise two
probabilistic states of a particle appear. The first one is described by the
probability density $\mathcal{P}(x,t)$ and is realized with probability
$\mathcal{P}_{0}(t)$. Since the noise does not affect the particle trajectory
$z(t)$, see (\ref{sol1}), we refer to this state as the surviving state. The
second state, which is associated with the  probability density $\mathcal{A}
(x,t)$ (we recall that $\mathcal{A} (x,t) \to 0$ as $\kappa \to 0$), is
realized with probability $\mathcal{A}_{0}(t)$. The transition to this state
implies that a particle jumps to infinity and it is excluded from future
consideration. It is therefore reasonable to call this state absorbing. To
avoid any confusion, we emphasize that the act of absorption is considered here
as the transition of a particle into the state in which the probability to find
this particle in any finite interval equals zero. Thus, super-heavy-tailed
noise plays the role of an absorbing medium characterized by the rate of
absorption $q$. Accordingly, the Langevin equation (\ref{Langevin}) driven by
this noise describes the overdamped motion of a particle in an absorbing
medium.

As equation (\ref{sol1}) shows, before absorption, which is random in time, the
motion is deterministic and occurs under a force field $f(x,t)$. The random
motion of a particle in an absorbing medium can also be described within the
Langevin approach. To this end, we represent $\xi(t)$ in the Langevin equation
(\ref{Langevin}) as a sum of two independent noises: (i) super-heavy-tailed
noise $\xi_{1}(t)$, which models an absorbing medium, and (ii) noise
$\xi_{2}(t)$, which induces the random motion of a particle. Calculations
similar to those described above lead to the result $\mathcal{P} (x,t) =
e^{-qt}W(x,t)$, where $W(x,t)$ denotes the normalized to unity probability
density associated with the random motion of a particle. In particular, if
$\xi_{2}(t)$ is Gaussian white noise then $W(x,t)$ is the solution of the
ordinary Fokker-Planck equation, and if $\xi_{2}(t)$ is L\'{e}vy stable noise
then $W(x,t)$ is the solution of the fractional Fokker-Planck equation. In
these cases, the Langevin equation describes Brownian motion and L\'{e}vy
flights, respectively, in an absorbing medium.

\section{Illustrative example}
\label{sec:Ill}

In order to verify numerically that super-heavy-tailed noise acts as an
absorbing medium, we first specify the super-heavy-tailed probability density
$p(y)$ by choosing
\begin{equation}
    p(y) = \frac{\ln s}{2(s+|y|)\ln^{2}(s+|y|)},
    \label{p2}
\end{equation}
where $s(>1)$ is the density parameter. This symmetric probability density is
unimodal with the peak located at $y=0$ and has no characteristic scale. The
behavior of $p(y)$ at $y \to 0^{+}$ is characterized by the conditions $p(0) =
1/(2s\ln s)$ and $dp(y)/dy |_{y\to 0^{+}} = -(1 + 0.5 \ln s)/(s\ln s)^{2}$, and
at $y\to \infty$ by the asymptotic formula $p(y) \sim \ln s/(2y\ln^{2}y)$ [this
implies that $h(y) \sim \ln s/(2\ln^{2}y)$]. For this probability density the
slowly varying function (\ref{Y1}), after calculating the inner integral, takes
the form
\begin{equation}
    Y(\lambda) = \frac{\ln s}{2}\int_{0}^{\infty} dx
    \frac{\sin x}{\ln(s+\lambda x)}.
    \label{Y3}
\end{equation}
To find the leading term of $Y(\lambda)$ at $\lambda \to \infty$, which
determines the rate of absorption (\ref{q1}), we use the identity
\begin{equation}
    \frac{1}{\ln v} = \int_{0}^{1} \frac{du}{v^{u}} + \frac{1}{v \ln v}
    \label{rel2}
\end{equation}
with $v=s+\lambda x$. Substitution of this identity into (\ref{Y3}) yields
\begin{eqnarray}
    Y(\lambda) &=& \frac{\ln s}{2}\int_{0}^{1}du \int_{0}^{\infty}
    dx \frac{\sin x}{(s+\lambda x)^{u}}
    \nonumber\\
    &&+ \frac{\ln s}{2}\int_{0}^{\infty} dx \frac{\sin x}
    {(s+\lambda x) \ln(s+\lambda x)}.
    \label{Y4}
\end{eqnarray}
Then, taking into account that $\int_{0}^{\infty}dx \sin(x)/x^{u} = \Gamma
(1-u) \sin[\pi(1-u)/2]$, where $\Gamma(x)$ is the gamma function,
\begin{eqnarray}
    \int_{0}^{1}du \int_{0}^{\infty} dx\frac{\sin x}{(s+\lambda x)
    ^{u}} &\sim& \int_{0}^{1}du \frac{\Gamma(1-u)}{\lambda^{u}}
    \sin \frac{\pi(1-u)}{2}
    \nonumber\\
    &\sim& \int_{0}^{1}du\, e^{-u\ln \lambda} \sim \frac{1}
    {\ln \lambda}  \quad\; (\lambda \to \infty)
    \label{int1}
\end{eqnarray}
and
\begin{equation}
    \int_{0}^{\infty} dx \frac{\sin x}{(s+\lambda x)
    \ln(s+\lambda x)} = o\! \left(\frac{1}{\lambda} \right)
    \quad\; (\lambda \to \infty),
    \label{int2}
\end{equation}
we arrive to the desired asymptotic formula $Y(\lambda) \sim \ln s \,(2\ln
\lambda)^{-1}$ ($\lambda \to \infty$).

As it follows from the definition of the rate of absorbtion (\ref{q1}), in the
considered case it can be represented in the form
\begin{equation}
    q = \frac{\ln s}{r}
    \label{q1a}
\end{equation}
with
\begin{equation}
    r = \lim_{\tau \to 0} \ln\! \left( \frac{1}{a(\tau)}
    \right)^{\!\tau}\!.
    \label{r}
\end{equation}
Depending on the behavior of the scale function $a(\tau)$ at $\tau \to 0$, the
parameter $r$ can take the values from $0$ to $\infty$. If $a(\tau)$ approaches
zero so fast that $r=\infty$ then super-heavy-tailed noise does not affect the
particle dynamics at all, i.e. the particle absorption is absent ($q=0$). In
contrast, if $a(\tau)$ tends to zero in such a way that $r=0$ then the noise
action is so strong that the transition to the absorbing state happens at
$t=0^{+}$, i.e. a particle is absorbed immediately ($q= \infty$). A non-trivial
action of super-heavy-tailed noise occurs at $0<r<\infty$. According to
(\ref{r}), in this case $a(\tau) \sim b(\tau) e^{-r/\tau}$, where $b(\tau)$ is
a positive function satisfying the condition $[b(\tau)]^{\tau} \to 1$ as $\tau
\to 0$. An explicit form of this function is of no importance (and so we can
choose $a(\tau) = e^{-r/\tau}$) because at $a(\tau) \sim b(\tau) e^{-r/\tau}$
and $\lim_{\tau \to 0}[b(\tau)]^{\tau} =1$ the rate of absorption $q$ does not
depend on $b(\tau)$:
\begin{equation}
    q= \frac{\ln s}{\lim_{\tau \to 0} \ln [e^{r/\tau}/b(\tau)]^{\tau}} =
    \frac{\ln s}{r - \lim_{\tau \to 0} \ln [b(\tau)]^{\tau}}=
    \frac{\ln s}{r}.
    \label{q2}
\end{equation}
Thus, assuming for simplicity that $f(x,t)=0$, we obtain for the examined case
the probability density of the particle position in the surviving
(non-absorbing) state,
\begin{equation}
    \mathcal{P}(x,t) = s^{-t/r}\delta(x),
    \label{pdf}
\end{equation}
and the probability of absorption, $\mathcal{A}_{0}(t) = 1 - s^{-t/r}$.

It is worthy to note that while the probability density (\ref{p2}) with
super-heavy tails depends on the single parameter $s$, the corresponding
super-heavy-tailed noise is characterized also by the additional parameter $r$.
The parameter $r$ relates to the scale function $a(\tau)$ and at a fixed $s$ it
controls the noise intensity, namely, the smaller the parameter $r$ is the
larger is this intensity. The dependence of the scale function on the parameter
which characterizes the noise intensity is not surprising and occurs for all
probability densities $p(y)$ \cite{DHH2}. If the parameters $s$ and $r$ are
chosen then super-heavy-tailed noise associated with the probability density
(\ref{p2}) is completely determined and both the Langevin and Fokker-Planck
equations are well defined.

\section{Numerical simulations}
\label{sec:Num}

We have verified the analytical results obtained in the previous section by
numerically studying the statistics of the particle positions. Because of the
Langevin equation has never before been simulated in the case of
super-heavy-tailed noises, we briefly describe the procedure for finding the
particle position $x(n\tau) = \sum_{j=1}^{n}\Delta \eta (j\tau)$ at $t=n\tau$.
The first step is calculating the quantity $\rho_{j} = a(\tau)(s^{1/(1-z_{j})}
-s)$ with $a(\tau) = e^{-r/\tau}$ and $z_{j}$ being a random number uniformly
distributed in the interval $[0,1]$. Since the positive quantities $\rho_{j}$
are distributed with the probability density $\ln s \, \{ a(\tau)[s + \rho
/a(\tau)] \ln^{2} [s + \rho /a(\tau)] \}^{-1}$, the increments $\Delta \eta(j
\tau)$, whose probability density $p(\Delta \eta/a(\tau))/ a(\tau)$ is
determined by (\ref{p2}), can be represented in the form $\Delta \eta(j \tau) =
(\pm)_{j}\, \rho_{j}/2$, where $(\pm)_{j}=+1$ or $-1$ with probability $1/2$
each. Finally, calculating $\Delta \eta(j \tau)$ $n$ times, we obtain the
simulated value of the particle position: $x(n\tau) = \sum_{j=1}^{n}
(\pm)_{j}\, \rho_{j}/2$.

The sample paths of $x(n\tau)$ are the piecewise constant random functions of
time which are defined on the intervals $[j\tau, j\tau + \tau)$ ($0\leq j\leq
n$) and satisfy the condition $x(0)=0$ (see figure~1(a)). The behavior of these
sample paths at $\tau \to 0$ is determined by two major factors. The first
consists in the absence of finite fractional moments of $\rho_{j}$. This means
that the increments $\Delta \eta (j\tau)$ have no characteristic scale and
arbitrary large values of $\Delta \eta (j\tau)$ are in principle possible. The
second is that the scale function $a(\tau)$ tends to zero very rapidly as
$\tau$ decreases. Since $\Delta \eta (j\tau) \propto a(\tau)$, this implies
that for small enough $\tau$ the difference between the increments in
neighboring intervals is in general negligible. Therefore, one may expect (and
this is confirmed by our simulations) that in the limit $\tau \to 0$ the
competition of these factors leads to the condition $x(t) = 0$ with
$t<t_{\rm{tr}}$, where $t_{\rm{tr}}$ is a random time in which the absolute
value of the increment becomes bigger than any preassigned positive number (see
figure~1(b)). We interpret the discontinuous jump of $x(t)$ which occurs at $t=
t_{\rm{tr}}$ as the transition of a particle to the absorbing state.
\begin{figure}[ht]
    \centering
    \includegraphics[totalheight=3.7cm]{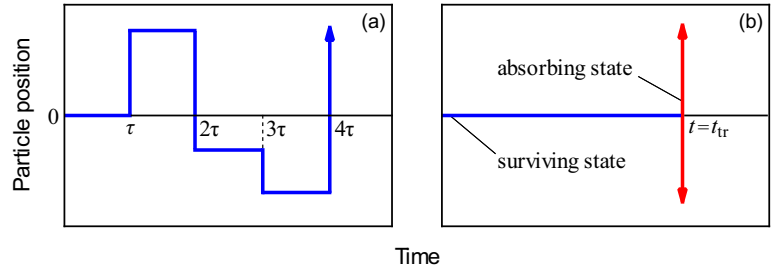}
    \caption{\label{fig1} Illustrative examples of the sample
    paths of the particle position $x(t)$ for a finite
    time increment $\tau$ (a) and in the limit $\tau \to 0$ (b).
    The vertical arrow in figure~1(a) means that the jump of
    $x(t)$ at $t=4\tau$ exceeds the figure frame. In contrast,
    the arrows in figure~1(b) indicate that at $t = t_{\rm{tr}}$
    an infinite jump of $x(t)$ occurs in the positive or negative
    direction of the axis $x$. The horizontal line with $x(t) =0$
    and $t<t_{\rm{tr}}$ corresponds to the surviving state, and
    the vertical line to the absorbing state.}
\end{figure}

Illustrative  histograms of the probability distribution of particle positions
are shown in figure~2 for two different values of the time increment $\tau$.
\begin{figure}[ht]
    \centering
    \includegraphics[totalheight=4cm]{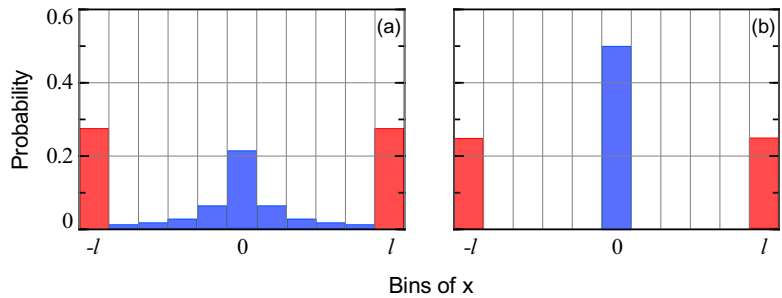}
    \caption{\label{fig2} Histograms of the probability
    distribution of particle positions obtained by numerical
    simulations of the Langevin equation (\ref{Langevin}).
    The parameters of the simulation are $t=1$, $r=1$, $s=2$,
    $N=10^{6}$, $l=5$, $\epsilon = 0.2$, and $\tau=0.25$ (a)
    and $\tau=10^{-3}$ (b).}
\end{figure}
The height of the $m$th bin ($m=0,\pm 1,\ldots, \pm(l-1)$) represents the
probability $W_{m}$ to find a particle in this bin at time $t=n\tau$. The
probability $W_{m}$ is determined as the ratio $N_{m}/N$, where $N$ is the
overall number of numerical experiments, i.e. calculations of $x(t)$, and
$N_{m}$ is the number of those experiments in which $x(t)$ falls into this bin,
i.e. when $x(t) \in (m\epsilon - \epsilon/2, m\epsilon + \epsilon/2)$ with
$\epsilon$ being the bin width. In contrast, the heights of the leftmost and
rightmost bins represent the probabilities $W_{-l}$ and $W_{l}$ that $x(t)<-l
\epsilon + \epsilon/2$ and $x(t)>l \epsilon - \epsilon/2$, respectively.
According to these definitions $\sum_{m=-l}^{l} W_{m} =1$. The probabilities
$W_{m}$ ($m\neq 0, \pm l$) rapidly tend to zero as $\tau$ decreases, and for
small enough $\tau$ the histogram takes the form with only three visible bins
(e.g., in figure~2(b) ${\rm{max}}\,W_{m} \sim 10^{-4}$, $W_{0} \approx 0.5$,
$W_{-l} \approx W_{l} \approx 0.25$). Hence, the less is $\tau$ the better the
condition $W_{0} + W_{-l} + W_{l} =1$ holds. The probability $W_{0}$ approaches
the survival probability $s^{-t/r}$ (see figure~3),
\begin{figure}
    \centering
    \includegraphics[totalheight=4cm]{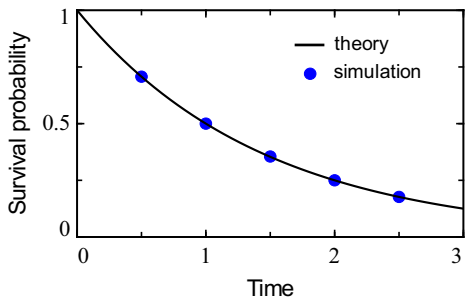}
    \caption{\label{fig3} Time dependence of the
    survival probability. The solid line represents the
    theoretical dependence $s^{-t/r}$ (for $r=1$, $s=2$),
    and the circular symbols indicate the results of the
    numerical simulation. The latter are obtained by calculating
    the height of the central bin in figure~1(b) for different
    values of the dimensionless time $t$.}
\end{figure}
and its independence on $\epsilon$ (if $\epsilon$ is not too small for a fixed
$\tau$) yields $W_{0} = \int_{-\epsilon}^{\epsilon}dx\,\mathcal{P}(x,t)$, i.e.
the theoretical result (\ref{pdf}) indeed holds. Finally, the facts that
$W_{m}$ ($m\neq 0, \pm l$) approach zero as $\tau$ decreases and $W_{-l} +
W_{l} \approx 1 - s^{-t/r}$ (if $l$ is not too large for a fixed $\tau$)
capture the main properties of $\mathcal{A} (x,t)$, namely, $\mathcal{A}(x,t)
\to 0$ as $\kappa \to 0$ and $\mathcal{A}_{0}(t) = 1 - s^{-t/r}$. Thus, the
simulation results are in full agreement with our theoretical predictions.

\section{Conclusion}
\label{sec:Concl}

In summary, we have incorporated a class of super-heavy-tailed noises into the
Langevin equation formalism. These noises arise from the super-heavy-tailed
probability densities whose all fractional moments are infinite. Due to this
feature, super-heavy-tailed noises induce two probabilistic states of a
particle, survived and absorbed ones. Thus the Langevin equation driven by
these noises becomes a useful tool for studying a number of physical phenomena
such as absorption, evaporation, annihilation, etc. We have solved analytically
the corresponding generalized Fokker-Planck equation and have confirmed our
theoretical predictions via the numerical simulation of this Langevin equation.

\section*{Acknowledgments}
\label{sec:Ack}

The authors are grateful to A~Yu~Polyakov for his help with the numerical
calculations.

\section*{References}
\label{sec:Ref}

\end{document}